\documentstyle[preprint,prl,aps]{revtex}
\begin{document}
\preprint{FTUAM/95-18}
\draft
\title{
Approximate particle number projection for finite range
density dependent forces}

\author{A. Valor, J.L. Egido, L.M. Robledo}
\address{ Departamento de F\'\i sica Te\'orica C-XI\\
        Universidad Aut\'onoma de Madrid, E--28049 Madrid, Spain }

\maketitle

\begin{abstract}

The Lipkin-Nogami method is generalized to deal with finite range
density dependent forces.  New expressions are derived and realistic
calculations with the Gogny force are performed for the nuclei
$^{164}$Er and $^{168}$Er.  The sharp phase transition predicted by
the mean field approximation is washed out by the Lipkin-Nogami
approach. A much better agreement with the experimental data is
reached with the new approach than with the Hartree-Fock-Bogoliubov
one, specially at high spins.

\end {abstract}

\narrowtext

Mean-field theories (BCS, Hartree-Fock and Hartree-Fock-Bogoliubov,
HFB) are the cornerstone of all microscopic approximations to the
nuclear many body problem.  The success of these approaches is mostly
due to their ability to deal with single particle motion as well as
with the collective motion associated with symmetries~\footnote{
Continuous symmetries as rotations in any space: coordinate space,
gauge space of particle number operator, etc, as well as with discrete
symmetries, e.g. spatial parity.}, the latter one by enlarging the
variational Hilbert space with wave functions that are not eigenstates
of the symmetry operators.  These wave functions are usually
constrained to accomplish the symmetries on the average; for most
symmetries (with the exception of the particle number) this is a very
satisfactory approach, see Ref.~\cite{RS.80} for a thorough
discussion.  In the case of pairing correlations, in which we are
interested in this letter, the crucial quantities are the number of
correlated pairs and the level density around the Fermi surface. If
these quantities are small, and in nuclei they usually are, mean field
theories are not enough and one should do something better.

The ideal treatment of pairing correlations in nuclei is particle
number projection before the variation \cite{RS.80}. At high spin this
theory is rather complicated and up to now it has only been applied
with separable forces \cite{EGPNP}.  On the other hand, the
semi-classic recipe of solving the mean field equations with a
constraint on the corresponding symmetry operator can be derived as
the first order result of a full quantum-mechanical expansion (the
Kamlah expansion) \cite{KAM.68} of the projected quantities.  The
second order in this expansion takes into account the particle number
fluctuations and might cure some of the deficiencies of the first
order approximation. However, full calculations up to second order are
rather cumbersome and just a simple model calculation has been carried
out up to now \cite{FLO.94}.  Most second order calculations have been
done using the Lipkin-Nogami (LN) recipe originally proposed in
Refs.~\cite{LIP.60,NOG.64,GN.66}, see also the clarifying papers by
Quentin et al. \cite{QUE.90} and by Flocard at al. \cite{FLO.94}.
Recently the LN method has been applied to study superdeformed nuclei
at high spins adding a monopole pairing interaction to the Woods-Saxon
plus Strutinsky \cite{MCD.93} and to the Skyrme force \cite{GBD.94}.

Up to now the LN method has been only formulated for the case of a
simple separable pairing interaction. The purpose of this Letter is to
extend such studies to more realistic pairing interactions like the
one implicit in the finite range and density dependent Gogny force
where the particle-hole and particle-particle part of the interaction
are generated from the same force.  We have formulated the
Lipkin-Nogami method using the Kamlah expansion and treated the
density dependence consistently. In the new formulation additional
terms arise in the equations determining the parameters of the theory.

Let $|\Phi \rangle$ be a product wave function of the
Hartree-Fock-Bogoliubov type, i.e. a particle number symmetry
violating wave function. We can generate an eigenstate of the particle
number $|\Psi_{N}\rangle$ by the projection technique \cite{RS.80}
\begin{equation}
   |\Psi_{N}\rangle = \hat{P}^{N}|\Phi \rangle=
   \frac{1}{2\pi} \int_{0}^{2\pi}d\phi
e^{i(\hat{N}- N )\phi}  |\Phi \rangle.
\end{equation}
The particle number projected energy is given by
\begin{eqnarray}
  E^{N}_{proj}  =  \frac{\langle\Psi_{N}|\hat{H}|\Psi_{N} \rangle}
  {\langle\Psi_{N}|\Psi_{N} \rangle}=
    \frac{
   \int_{0}^{2\pi}d\phi e^{-i\phi N}\langle\Phi|\hat{H}e^{i\phi\hat{N}}
   |\Phi\rangle}
   {\int_{0}^{2\pi}d\phi e^{-i\phi N}\langle\Phi|
   e^{i\phi\hat{N}} |\Phi\rangle} =
    \frac{ \int_{0}^{2\pi}d\phi e^{-i\phi N}h(\phi)}
   {\int_{0}^{2\pi}d\phi e^{-i\phi N}n(\phi)}
 \label{bigeq}
\end{eqnarray}
where we have introduced the hamiltonian-, $h(\phi)=
\langle\Phi|\hat{H}e^{i\phi\hat{N}}|\Phi\rangle$,
and norm-, $n(\phi)=\langle\Phi|e^{i\phi\hat{N}}|\Phi\rangle$,
overlaps.  In the case of large particle numbers and strong
deformations in the gauge space associated to $\hat{N}$, one expects
the $ h(\phi)$ and $n(\phi)$ overlaps to be peaked at $\phi=0$ and to
be very small elsewhere in such a way that the quotient $
h(\phi)/n(\phi)$ behaves smoothly.  One can make an expansion of $
h(\phi)$ in terms of $n(\phi)$ in the following way \cite{KAM.68}
\begin{equation}
h(\phi)=\sum_{m=0}^{M}h_{m}{\hat{\cal{N}}}^{m}n(\phi)
\label{hphi}
\end{equation}
where we have introduced the Kamlah operator
 $ \hat{\cal{N}}= \frac{1}{i}\frac{\partial}
 {\partial{\phi}}-\langle\Phi|\hat{N}|\Phi\rangle$
which is a representation of the particle number operator
in the space of the parameter $\phi$.
The expansion coefficients $h_m$ are determined  by
applying the operators
 $1, \hat{\cal{N}},..., \hat{\cal{N}}^M$ on Eq.~(\ref{hphi}) and
taking the limit $\phi \rightarrow 0$.
{}From now on we shall use the shorthand notation
 $\langle {\hat A} \rangle=
 \langle \Phi | {\hat A}|\Phi \rangle $ and
 $\Delta\hat{N} = \hat{N} - \langle \hat{N} \rangle$.

The projected energy to second order is
\begin{equation}
 E_{proj}^{(2)}\;=\;\langle\hat{H}\rangle\;\;-\;\;
 h_{2}\langle(\Delta\hat{N})^{2}\rangle
 \;\;+\;\; h_{1}\;(N - \langle\hat{N}\rangle) \;\;+\;\;
 h_{2}\; (N - \langle\hat{N}\rangle)^{2},
\label{energy}
\end{equation}
with $h_1$ and $h_2$ given by
\begin{eqnarray}
  h_{1} & = & \frac{\langle\hat{H}\Delta\hat{N}\rangle -
  h_{2} \;\langle(\Delta\hat{N})^{3}\rangle}
 {\langle(\Delta\hat{N})^{2}\rangle}  \label{h1}\\
  h_{2} & = & \frac{\langle(\hat{H}-\langle\hat{H}\rangle)
  (\Delta\hat{N})^{2}\rangle  -
 \langle\hat{H}\Delta\hat{N}\rangle\langle(\Delta\hat{N})^{3}\rangle/
 \langle(\Delta\hat{N})^{2}\rangle}
 {\langle(\Delta\hat{N})^{4}\rangle - \langle(\Delta\hat{N})^{2}
 \rangle^{2}- \langle(\Delta\hat{N})^{3}\rangle^{2}/\langle
 (\Delta\hat{N})^{2}\rangle}.
\end{eqnarray}
In a full variation after projection method one should vary
Eq.~(\ref{energy}).  In the Lipkin-Nogami prescription, however, the
coefficient $h_2$ is held constant during the variation; the resulting
equation is much simpler, one gets
\begin{equation}
 \frac{\delta}{\delta \Phi}\langle \hat{H} - h_2 (\Delta\hat{N})^{2}\rangle
 - h_1 \frac{\delta}{\delta \Phi}\langle \hat{N}\rangle = 0,
\label{cran}
\end{equation}
with $h_1$ determined by  the constraint
\begin{equation}
       \langle \hat{N} \rangle = N.
\label{constr}
\end{equation}
If there are additional constraints, for example the angular momentum,
one just has to substitute $\hat{H}$ by $\hat{H^\prime}=\hat{H}-
\omega \hat{J_x}$ in all equations above and to add the constraint
$\langle \hat{J_x} \rangle = [I(I+1)-\langle \hat{J_z}^2 \rangle
]^{1/2} $ to the one of Eq.~(\ref{constr}).

Now we would like to generalize the formulae above to density
dependent forces, like the Gogny force \cite{GOG.75,BGG.84} which has
a term proportional to $\rho^{\alpha}(\frac{\vec{r}_1+\vec{r}_2}{2})$.

In this case the density term\footnote{The dependence of the density
with $\phi$ is given by $\rho (r) = \langle\Phi | c^\dagger (r) c (r)
e^{i\phi\hat{N}} |\Phi \rangle /\langle\Phi| e^{i\phi\hat{N}}|\Phi
\rangle$.} causes a dependence on $\phi$ of the hamiltonian
\cite{BDF.90}.  The Kamlah expansion (\ref{hphi}) provides in this
case the following equation system
\begin{eqnarray}
\label{meq4fin}
\left.\left\langle \frac{1}{i}\frac{\partial\hat{H}}
{\partial\phi}\right|_{\phi=0} +
 \hat{H}\Delta\hat{N}\right\rangle & = &
     h_1\langle(\Delta\hat{N})^2\rangle + h_2\langle
     (\Delta\hat{N})^3\rangle \\
\left\langle \left.\left. \frac{1}{i^2}\frac{\partial^{2}
\hat{H}}{\partial\phi^2}
\right|_{\phi=0} + 2\Delta\hat{N}
\frac{1}{i}\frac{\partial\hat{H}}{\partial\phi}\right|_{\phi=0}
\; +\; \hat{H}(\Delta\hat{N})^{2}\right\rangle & = &
       \langle H \rangle \langle(\Delta\hat{N})^{2}\rangle \nonumber \\
   & + &  h_1\langle(\Delta\hat{N})^3\rangle
     + h_2 \left( \langle(\Delta\hat{N})^4\rangle -
     \langle(\Delta \hat{N})^2\rangle^2 \right)
\nonumber
\end{eqnarray}
which determine the coefficients $h_1$ and $h_2$.

{}From now on we proceed as in the non-density dependent case, i.e. we
have to solve Eq.~(~\ref{cran}~) with the constraint (\ref{constr})
but with the coefficient $h_1$ and $h_2$ provided by the equation
system (\ref{meq4fin}).

We have applied this formalism to study high spin states with the
Gogny force in two rare earth nuclei: $^{164}$Er as an example of
strong back-bender and $^{168}$Er as a non-back-bender. We shall refer
to this calculations as cranked-HFB-LN (CHFBLN).  In all our results
we shall also present the ones obtained with the plain cranked HFB
theory (CHFB) \cite{ER.93}.  We are using the standard DS1
parametrization set \cite{BGG.84}. Of course, one could ask if the
parametrization of the force should be changed, for it was adjusted
for plain mean field calculation. Since this is the first
investigation in this direction we shall use the standard
parametrization, further investigations will be reported in the
future.

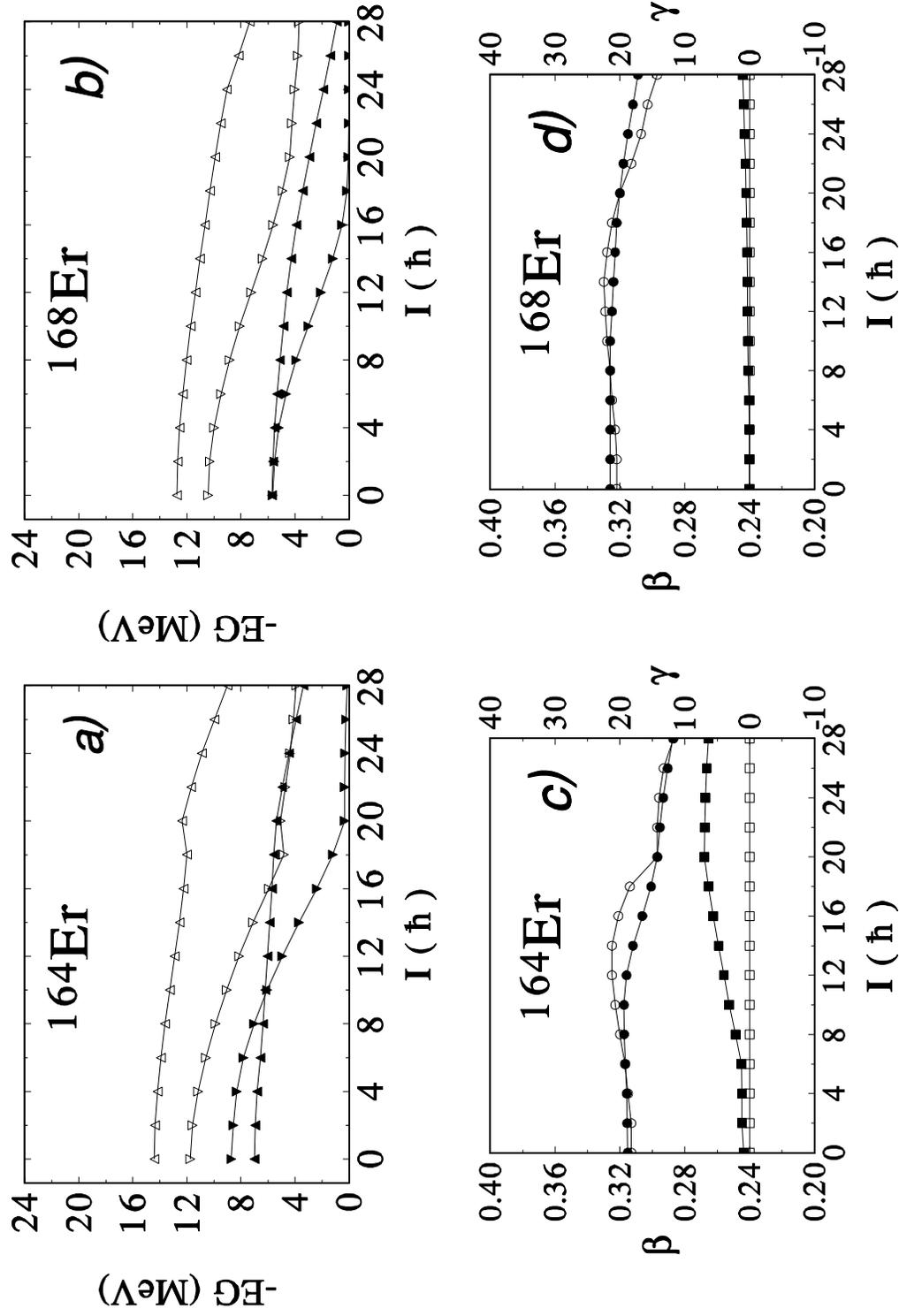
\begin{figure}
\caption{ Upper panels:  Pairing energies versus the angular momentum.
Proton (neutron) values are represented by triangles (inverted
triangles). Lower panels: Deformation parameters $\beta$ (circles,
scale on the left axis) and $\gamma$ (squares, scale on the right
axis) versus angular momentum.  Full (empty) symbols correspond to
CHFB (CHFBLN).}
\label{fig1}
\end{figure}

In Fig.~\ref{fig1}a we show the pairing energy in the CHFB approach
and in the CHFBLN approach versus the angular momentum. For $^{164}$Er
in the CHFB approach we observe for the neutron system, first, a
strong Coriolis antipairing effect which diminishes the neutron
pairing correlations, later on the crossing of the ground state band
with a two neutron aligned band -see below- causes the quenching of
the pairing correlations at $I\simeq 18\hbar$.  The proton system, on
the other hand, behaves very smoothly until $I=18 \hbar$. From
$I=18\hbar$ up to $I=28\hbar$ we observe the typical Coriolis
antipairing effect reduction, which is not as strong as for the
neutron system because the intruder orbital in this case is a
$h_{\frac{11}{2}}$ at variance with the $i_{\frac{13}{2}}$ of the
neutrons. In the CHFBLN results the same Coriolis antipairing effects
are observed but no superfluid to normal fluid phase transition is
found. We also realize that the LN term has a larger effect on the
proton system than on the neutron one.  For $^{168}$Er,
Fig.~\ref{fig1}b, again the neutron phase transition is washed out in
the CHFBLN approach and a larger increase in the pairing energies of
the proton system than in the neutron one is obtained; this may have
to do with the different intruders for both systems.

The most relevant deformation parameters are $\beta$ and $ \gamma$ (we
define them as in Ref.~\cite{GG.83}). Their angular momentum
dependence is displayed in Figs.~\ref{fig1}c and \ref{fig1}d. For
$^{164}$Er, in the CHFB approach we first observe a rather constant
value of the deformation parameters $\beta$ until $I=12 \hbar$, from
this point on and until $I=28 \hbar$ we find a decrease in $\beta$.
This anti-stretching effect is caused by the Coriolis force. The
CHFBLN approach differs from the CHFB one in the spin range $I=
10\hbar$ till $I= 18\hbar$, where the neutron pairing collapse take
place. Along the aligned band the $\beta$ values are again very
similar in both approaches.  In the CHFB approach, at $I=0\hbar$ the
nucleus $^{164}$Er is axially symmetric ($\gamma=0$), then $\gamma$
increases up to 8 degrees at $I=18 \hbar$, later on it decreases very
slowly.  In the CHFBLN approach, the nucleus remains axially symmetric
($\gamma=0$) at all spin values.  In the nucleus $^{168}$Er we observe
a similar behavior when we compare the $\beta$-values of the CHFBLN
and the CHFB at the spin range $I=12-18\hbar$.   From $I=22-28\hbar$
we find a larger decrease in the $\beta$-values of the CHFBLN as
compared with the CHFB.  The reason for this behavior can be found in
Fig.~\ref{fig3}d; due to the smaller value of the moment of inertia in
the CHFBLN approach as compared with the CHFB one, larger values of
the cranking frequency are needed to produce the same angular
momentum.  These larger values of the cranking frequency causing an
stronger anti-stretching effect on the $\beta$-values of the CHFBLN
approach. The $\gamma$-values of $^{168}$Er are close to zero in both
approaches.

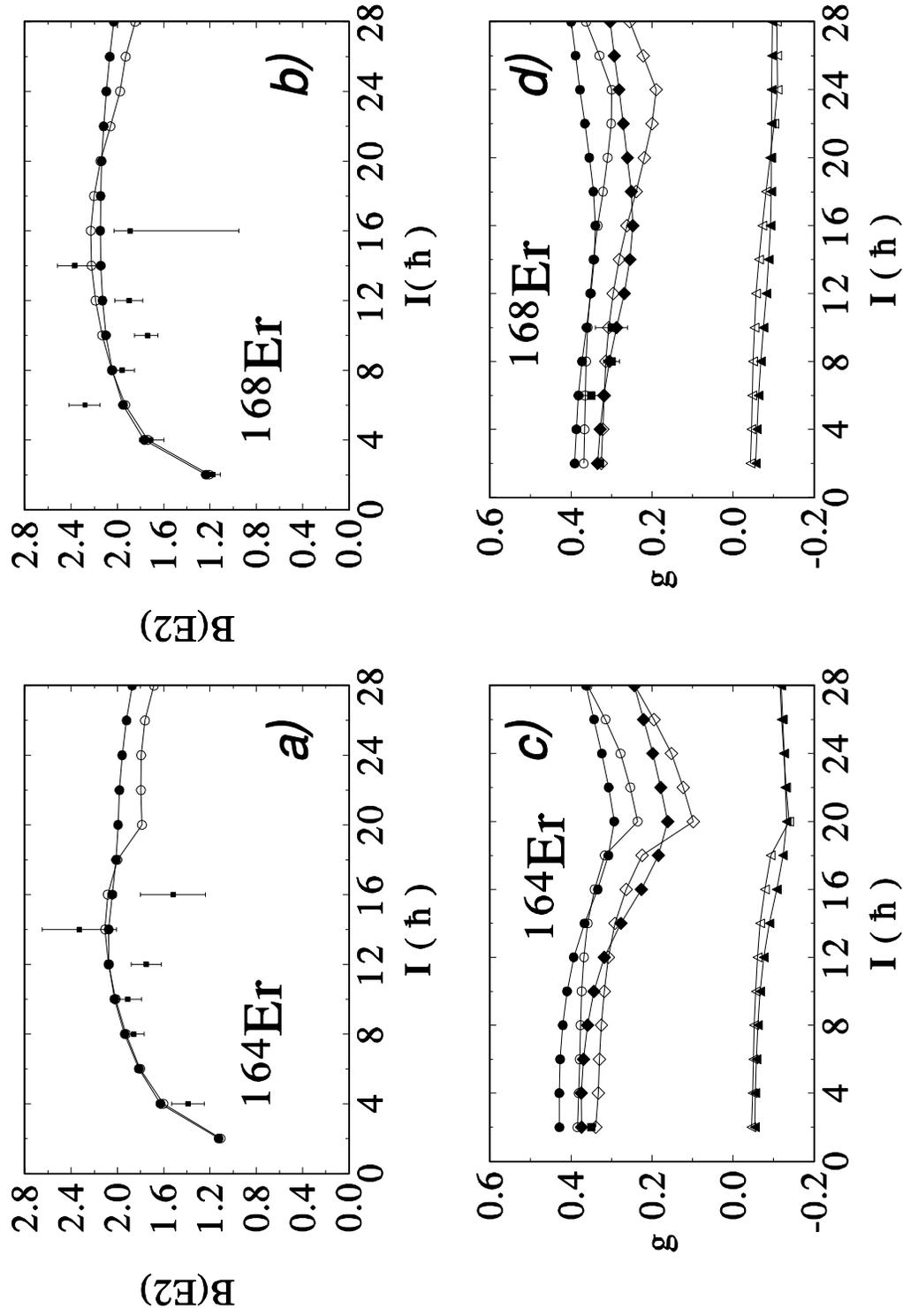
\begin{figure}
\caption{Upper panels: Reduced transition probabilities $B(E2)$
(in units of $(eb)^2$) along the Yrast band as a function of the
angular momentum. Full (open) circles stand for the CHFB (CHFBLN)
results and full squares for the experimental data (\protect
\cite{BE2EXP} for $^{164}$Er and
\protect \cite{BE2EXP8} for $^{168}$Er).
Lower panels: Gyromagnetic factors (in units of $\mu_N$ of
the Yrast states versus the  angular momentum; $g$ is
represented by diamonds, $g_p$ by  circles  and $g_n$ by
triangles. Full (open) symbols correspond to CHFB (CHFBLN).
The experimental data (\protect \cite{GEXP} for $^{164}$Er and
\protect \cite{GEXP8} for $^{168}$Er) are represented by full squares.}
\label{fig2}
\end{figure}

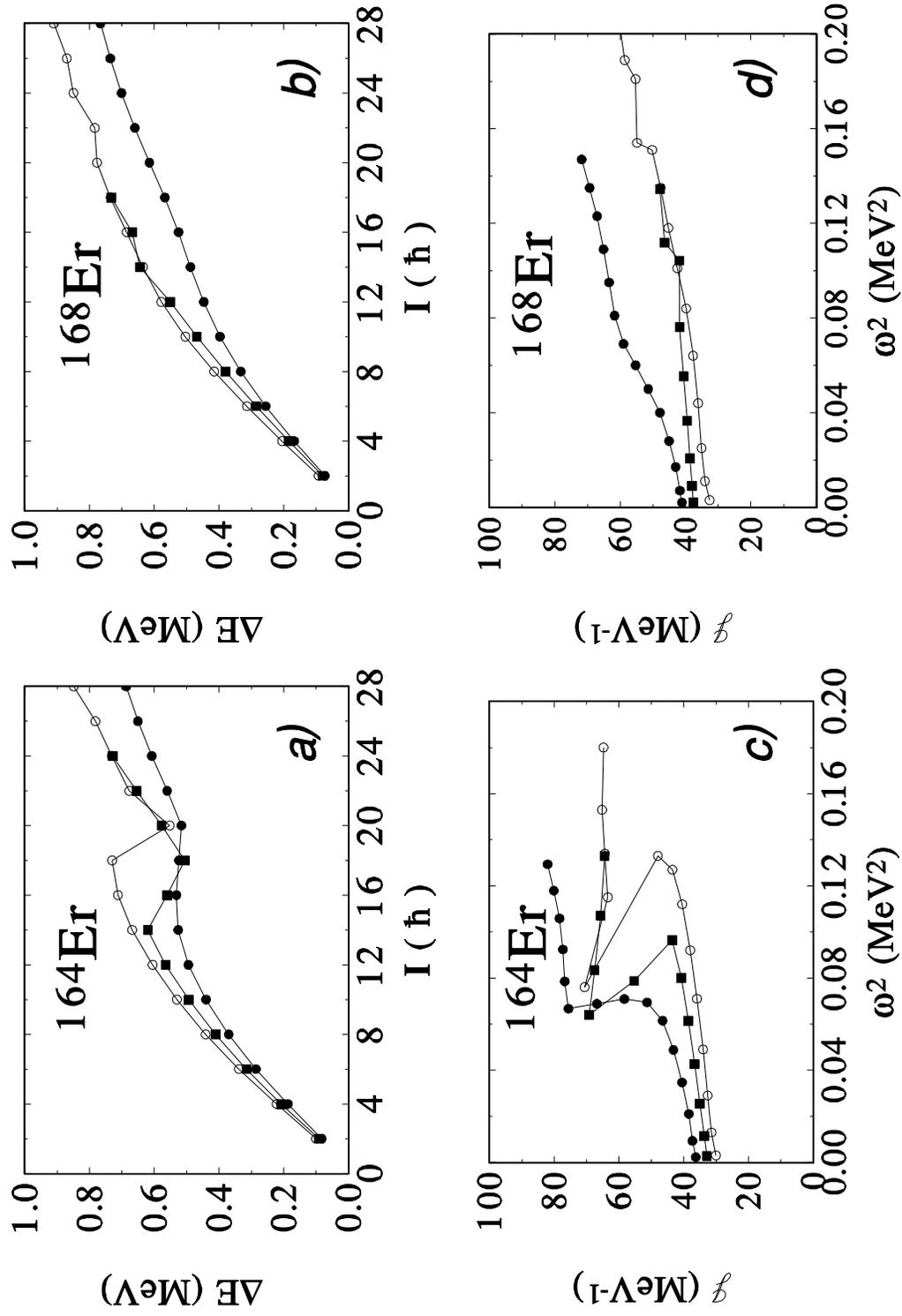
\begin{figure}
\caption{Upper panels: the gamma-ray energy $\protect \Delta
E_I=E(I)-E(I-2)$ as a function of the angular momentum.
Full (open) circles stand for the CHFB (CHFBLN) results and full
squares  for the experimental data (\protect \cite{BE2EXP}
for $^{164}$Er and
\protect \cite{BE2EXP8} for $^{168}$Er).
Lower panels: the  moment of inertia
$\protect {\cal J}= (2I-1)/\Delta E_I$
versus the square of the angular frequency.
The meaning of the symbols is the
same as in the upper panels.}
\label{fig3}
\end{figure}

The $E2$ transition probabilities and the gyromagnetic factors have
been calculated in the cranking approximation \cite{ER.93}. Since our
configuration space is large enough (11 oscillator shells) no
effective charges have been used in the calculations.  In
Fig.~\ref{fig2}a we show the reduced transition probabilities along
the Yrast band versus the angular momentum and the experimental ones
for the nucleus $^{164}$Er. At spin values up to $I=10 \hbar$ our
theoretical results are in good agreement with the experimental data.
For spin values 12, 14 and $16 \hbar$, corresponding to the band
crossing, we are not able to reproduce the zig-zag behavior of the
experimental data. This result is not surprising since we know that
the cranking approximation is not good in the band crossing. The
decrease of the CHFBLN results as compared with the CHFB ones is due
to the fact that in the CHFBLN approach the nucleus remain axially
symmetric at this spin values while in the CHFB approach it does not.
Concerning $^{168}$Er, Fig.~\ref{fig2}b, neither CHFB nor CHFBLN are
able to reproduce the zig-zag at medium spins (this behavior is not
understood, to our knowledge, in any theory). The smaller values of
the CHFBLN at high spins are due to the smaller $\beta$-values of this
approach.

To investigate the alignment processes in these nuclei we shall study
the gyromagnetic factors. In the calculation the free values of the
orbital and spin gyromagnetic factors have been used and no rotor
contribution $g_R$ has been considered.  In Fig.~\ref{fig2}c we
display the theoretical $g-$factors $g,g_p$ and $g_n$ as well as the
experimental $g-$factor at $I=2\hbar$\cite{GEXP} for $^{164}$Er. From
the pattern of $g$ we can conclude that at low spins we have a smooth
neutron alignment, at medium spins and up to $I=18 \hbar$ we have
strong neutron alignment and for higher spins we observe proton
alignment. The tendency is qualitatively the same in both
approximations, the CHFBLN displaying a sharper behavior though.  For
$^{168}$Er, Fig.~\ref{fig2}d, we obtain in the CHFB (CHFBLN) an smooth
neutron alignment up to spin $16\hbar$ ($24\hbar$), later on proton
alignment.  The agreement with the known experimental data ($I=6,8$
and $10\hbar$) is excellent.

In Fig.~\ref{fig3}a we display transition energies versus the angular
momentum. For $^{164}$Er the agreement with the experimental data is
good at low spins in the CHFB approach while in the band crossing
region we see that the crossing is not as sharp as in the experiment
-this is a well known drawback of the cranking approximation. In the
high spin part we get smaller values for the transition energies than
in the experiment.  In the CHFBLN approach at low and medium spins the
agreement with the experiment is better than in the CHFB approach. In
the backbending region the results are not good again, but in the very
high spin region the agreement with the experiment is very good at
variance with the CHFB approximation.  Concerning $^{168}$Er,
Fig.~\ref{fig3}b, the CHFB results are very low as compared with the
experiment; the CHFBLN ones, however, are in excellent agreement with
the experiment.

In Fig.~\ref{fig3}c,d we display the moments of inertia, versus the
square of the angular frequency.  In $^{164}$Er in the CHFBLN approach
we obtain at low spins a smaller value than in the CHFB due to the
larger pairing correlations in better agreement with the experiment.
In the band crossing we obtain a clear back-bending although shifted
in a few units as compared with the experiment indicating, perhaps,
that angular momentum projection is important in this region.  For
spin values $I=18, 20, 22 \hbar$, i.e. on the aligned band, the
agreement with the experiment is excellent~\footnote{ Notice that the
experimental results end at spin $I=22\hbar$ ($I=16\hbar$) for
$^{164}$ Er ($^{168}$Er) while the theoretical ones go up to
$I=26\hbar$.}, at variance with the simpler CHFB approach.  Concerning
$^{168}$Er, Fig.~\ref{fig3}d, the results of the CHFBLN approach are
much better than the ones of CHFB, we are at low spins somewhat lower
than in the experiment, but at medium and high spins the agreement is
very good. We would like to stress that no ${\cal J}_c$, the core
moment of inertia, has been assumed.

In conclusion, for the first time, we have formulated the
Lipkin-Nogami approximation for density dependent forces. We have
performed numerical calculations with the finite range density
dependent Gogny force for two nuclei, the theoretical results with
this approximation are in much better agreement with the experiment
than the plain HFB calculations.

This work was supported in part by DGICyT, Spain under Project
PB91--0006.  One of us (A.V.) would like to thank the Spanish
Ministerio de Asuntos Exteriores for financial support through an ICI
grant.

\end{document}